\documentstyle[12pt,a4,amstex,ctagsplt,righttag]{article}
\newcommand\R{\Bbb{R}}

\newcommand\Z{\Bbb{Z}}
\newcommand\C{\Bbb{C}}

\newcommand\K{\Bbb{K}}
\newcommand\N{\Bbb{N}}
\newcommand\dt[1]{\frac{d #1}{d t}}

\newcommand\cpoly{\C \left[x_1,\ldots,x_n\right]}

\newcommand\kpoly{\K \left[x_1,\ldots,x_n\right]}
\newcommand\crat{\C \left(x_1,\ldots,x_n\right)}

\newcommand\krat{\K \left(x_1,\ldots,x_n\right)}

\newcommand\myref[1]{(\ref{#1})}

\newcommand{\polyc}{\C \left[x_1,x_2,x_3\right]}

\newtheorem{twier}{Theorem} 
\newtheorem{lemma}{Lemma}
\newenvironment{proof}{\noindent{\bf
Proof.}}{$\Box$\\[\medskipamount]} 
\title {On the algebraic non-integrability of the Halphen
system}
\author{
{\sc Andrzej J.~Maciejewski}\\
Institute of Astronomy, N. Copernicus University\\
Chopina 12-18, 87-100 Toru\'n, Poland\\
(e-mail: maciejka@@astri.uni.torun.pl ) \\[\medskipamount]
and\\[\medskipamount]
{\sc Jean-Marie Strelcyn}\\
D\'epartement de Math\'ematiques, Universit\'e de Rouen,\\
76821 Mont Saint Aignan Cedex, France, URA CNRS 1378\\
(e-mail: strelcyn@@univ-rouen.fr)\\
and \\
Laboratoire Analyse, G\'eom\'etrie et Applications, URA
CNRS 742,\\
Institut Galil\'ee, D\'epartement de Math\'ematiques,\\
Avenue J.-B.Cl\'ement,  93430 Villetaneuse, France\\
(e-mail: strelcyn@@math.univ-paris13.fr) }
\begin{document}
\maketitle
\begin{abstract}
It is proved that the Halphen system of ordinary
differential equations has
no non-trivial rational first integrals.
\end{abstract}
\section{Introduction}
different classes  of
systems of ordinary differential equations (in
abbreviation ODE) is again of
great actuality. In \cite{Moulin:95::} an algebraic method
of proving the
non-integrability of polynomial  systems of ODE with
homogeneous  right
sides of the same degree  was presented and tested on some
non-trivial
examples (see also \cite{Nowicki:94::a} and
\cite{Nowicki:94::} for  another applications). This
method, according

to the best of our knowledge,  was  presented for the
first time in the
fundamental book of J.-P.~Jouanolou
\cite{Jouanolou:79::}.  Indeed, on
pages 193--195 of \cite{Jouanolou:79::}, M.H.A.~Levelt,
the  referee of the
book, proved using this method that the Jouanolou system
of ODEs %
\begin{equation*}
\dt{x} = z^s,\qquad
\dt{y} = x^s,\qquad
\dt{z} = y^s;\qquad s\in \N,\quad s\geq 2,
\end{equation*}
definition).  In
fact, the basic ideas of the method were already
introduced by
M.N.~Lagutinskii in his pioneering, but, unfortunately,
completely unknown
works \cite{Lagutinskii:1908::,Lagutinskii:1911::}.  See
\cite{Dobrowolskii:93::,Dobrowolskii:94::}, where one can
find more details
on M.N.~Lagutinskii and his works on integrability which
are direct
continuation of the seminal Darboux paper
\cite{Darboux:1878::}.

In this note, we study the Halphen system of ODEs
\cite{Halphen:1881::a,Halphen:1881::b} defined on $\C^3$
(or $\R^3$): %
\begin{equation}
\label{halphen}
\left.
\begin{split}
\frac{dx_1}{dt}&=x_2x_3-x_1(x_2+x_3),\\
\frac{dx_2}{dt}&=x_3x_1-x_2(x_3+x_1),\\
\frac{dx_3}{dt}&=x_1x_2-x_3(x_1+x_2),
\end{split}
\qquad\right\}
\end{equation}
\begin{twier}
\label{ratint}
The Halphen system \myref{halphen} does not admit any
non-trivial rational
first integral.
\end{twier}
To do this we first apply the Lagutinskii-Levelt procedure
that allows to
state non-existence of a polynomial first integral.  To
finish the proof it
was necessary to supplement the method by more subtle
investigations where
specific properties of the Halphen system are crucial.

It is well known that for the polynomial systems of ODE
the non existence of
a non-trivial rational first integral is equivalent to the
non-existence of
an algebraic first integral. This is a consequence of
Bruns theorem
\cite[vol.III, chap. XVII]{Forsyth:59::},
\cite{Kummer:90::}. For

polynomial systems with the homogeneous right sides of the
same degree,

called  {\em the homogeneous \/ } systems, the
non-existence of a

rational first integral is equivalent to the non-existence
of
a meromorphic first integral \cite{Ziglin:82::b}.
Consequently,

the Halphen system \myref{halphen} does not admit any
non-trivial

algebraic or meromorphic first integral.

The Halphen system appears in different contexts, e.g., as
a result of the
self-dual Yang-Mills reduction  \cite{Chakravarty:93::},
\cite{Takhtajan:92::}, \cite{Takhtajan:93::}, or, in
general relativity,

in the study of $SU(2)$-invariant four metrics (Bianchi IX
metrics)
\cite{Strachan:94::}, \cite{Chakravarty:94::}.

It is worth noticing that, in spite of the lack of
meromorphic first
integrals, the Halphen system can be explicitly
integrated---we can express
its general solution in terms of  elliptic integrals with
`variable modulus'
(see \cite{Halphen:1881::a}, \cite{Takhtajan:92::},
\cite{Strachan:94::}).
Moreover, as it was mentioned in \cite{Takhtajan:93::} and
\cite{Takhtajan:94::}, the  first integrals  of the
Halphen system do

indeed exist, although they are not global and are
multi-valued
non-algebraic functions. See also \cite{Grumal:93::} for
discussion of
special properties of the Halphen system.  %

The paper is organized as follows. In Sec.~2 we gather all
necessary
algebraic facts and present the basic steps of the
Lagutinskii-Levelt
procedure. The proof of our theorem is given in Sec.~3.  %
\section{Algebraic preliminaries}
\begin{equation}
\label{gensys}
\dt{x_i}=V_i(x_1, \ldots, x_n),  \qquad 1\leq i\leq n,
\end{equation}
Here,
as usual, we denote by $\kpoly$ the polynomial ring in $n$
variables
${x_i}$, ${1\leq i\leq n}$,  with coefficients in a
commutative field $\K$,
and by $\krat$ the field of rational fractions of $n$
variables with
coefficients from $\K$. Throughout this note we assume
that $\K=\C$.  %

A function $F$ is a {\em first integral} of  system
\myref{gensys} if it
satisfies the following equation %
\[
{d_V}(F)\stackrel{\mathrm{def}}{=}\sum_{i=1}^n
V_i\partial_i F=0,\qquad
\mbox{where}\quad \partial_i F=\frac{\partial F}{\partial
x_i}.
\]
notion of derivations
is used. A derivation $d$ is a linear mapping of $\kpoly$
(or of $\krat$)
into itself satisfying the Leibnitz rule
$d_V(FG)=Gd_V(F)+Fd_V(G)$.  Thus,
we can  talk about derivation $d_V$ instead  of  system
\myref{gensys}. The
derivation $d_V$ is called {\em homogeneous} if the
corresponding system of
ODE \myref{gensys} is homogeneous.

The main object in our investigation are  the so called
{\em  Darboux
polynomials\/}  of a derivation $d_V$ (or {\em partial
first integrals\/} of
a system of ODEs \myref{gensys}), i.e.,  polynomials $F\in
\cpoly\backslash\C$ such that %
\begin{equation*}
{d_V}(F)= \sum_{i=1}^n V_i\partial_i F=PF,
\end{equation*}
Darboux polynomial $F$
is nothing else but a first integral of the system of ODEs
called also a
{\em  constant of the derivation $d_V$}.  These
polynomials, as an
investigation tool   of the integrability of the system
\myref{gensys},  were introduced for the first time by
Darboux in \cite{Darboux:1878::}.

It is easy to prove the following facts (see
\cite{Moulin:95::}).
\begin{enumerate}
\item
  An element $F=A/B\in\crat$,  with relatively prime
polynomials $A, B\in
\cpoly$ is a first integral of \myref{gensys} if and only
if $A$ and $B$ are
Darboux polynomials with the same `eigenvalue' $P$,  i.e.,
$d_V(A)=PA$, and
$d_V(B)=PB$.%
\item
 If $F\in\cpoly$ is a Darboux polynomial of
\myref{gensys}, then all its
irreducible factors are also Darboux polynomials.%
\item
The finite product of Darboux polynomials is also a
Darboux polynomial.
More precisely,
\[
 \text{if}\quad d_V(F_i)=P_iF_i, \quad 1\leq i \leq s,
\qquad
 \text{then}\quad
{d_V}(\prod_{i=1}^sF_i)=(\sum_{i=1}^sP_i)(\prod_{i=1}^sF_i)
{}.
\]
\item
For a homogeneous derivation $d_V$ (or a homogeneous
system) if
${d_V}(F)=PF$ for some $F,P\in\cpoly$ then also
${d_V}(F^+)=P^+F^+$, where
by $G^+$ we denote the homogeneous component of the
highest degree of a
polynomial $G$.
\item
For  homogeneous derivations $d_V$ (or homogeneous
systems), the following
assertion holds \cite[Lemma 2.1]{Moulin:95::}.  If $F$ is
a Darboux
polynomial, with  eigenvalue $P$ such that ${d_V}(F)=PF$,
then $P$ is a
homogeneous polynomial, and all homogeneous components of
$F$ are also
Darboux polynomials corresponding to the same $P$.
\item

\label{nomore}
Let us assume that $F_i$, for $1\leq i\leq s$, are all (up
to a
multiplicative constant) irreducible  homogeneous Darboux
polynomials of  a
homogeneous  derivation $d_V$, and  let $d_V(F_i)=P_iF_i$,
for $1\leq i\leq
s$.  In such a case if the polynomials $P_i$, for $1\leq
i\leq s $, are
linearly independent over $\Z$, then any Darboux
polynomial of $d_V$ is (up
to a multiplicative constant) of the form
$\prod_{i=1}^s{F_i}^{\alpha_i}$,
where $\alpha_i, 1\leq i\leq s$, are non negative
integers. This assertion
easily follows from properties 2, 3 and 5 given above.
\end{enumerate}

Thus, the non-existence of non constant Darboux
polynomials implies the
non-existence  of  rational first of integrals of the
corresponding system
of ODEs. However, there  exist  systems with Darboux
polynomials that have
no rational first integral. We will show that the Halphen
system belongs to
this class of systems.

Now, let us present the basic steps of the
Lagutinskii-Levelt procedure.
Let us consider the system \myref{gensys} with
homogeneous right hand sides
of the same degree $k$.  To prove the non-existence of a
polynomial first
integral or Darboux polynomial we make use of {\em Darboux
points}, i.e.,
the points $z=(z_1,\ldots,z_n)\neq(0,\ldots,0)$ satisfying
the following
equations
\[
    V_i(z)= \lambda z_i, \qquad 1\leq i\leq
n,\qquad\text{for some}\quad
\lambda\in\C.
\]
 Note that the existence of a Darboux point $z$ for  the
system  is
equivalent to the existence of the straight line solution
of the form
\[
  x_i(t)=z_i \phi(t), \quad 1\leq i\leq n,\qquad
\text{where}\quad
\dt{\phi}=\lambda\phi^k.
\]

 Now, let us assume that a homogeneous polynomial $F$ of
degree $m\geq 1$ is
a Darboux polynomial of our system:
\begin{equation*}
  \sum_{i=1}^{n} V_i\partial_iF = P F.
\end{equation*}
Combining the above equation with the Euler identity \[
\sum_{i=1}^{n}
x_i\partial_iF = m F,
\]
we can eliminate one partial derivative of F, e.g.,
$\partial_n F$, and we
obtain the equation
\begin{equation}
\label{local}
  \sum_{i=1}^{n-1} (V_ix_n-V_nx_i)\partial_iF = (x_nP - m
V_n)F.
\end{equation}
Without any loss of generality, we can assume that the
last component of a
chosen Darboux point $z$ does not vanish, and we set
$z_n=1$.  Putting
$x_n=1$ in the equation \myref{local} and shifting  the
origin to the
Darboux point by the transformation
\[
  x_i = y_i + z_i, \qquad 1\leq i \leq n-1,
\]
we obtain the following equation
\begin{equation}
\label{reduced}
  \sum_{i=1}^{n-1} w_i\partial_i f = q f,
\end{equation}
where
\begin{equation}
\label{new}
\left.
 \begin{gathered} w_i=w_i(y_1, \ldots, y_{n-1}) = v_i(y_1,
\ldots,
y_{n-1})-v_n(y_1, \ldots, y_{n-1})(y_i + z_i),\\ q =
p(y_1, \ldots,
y_{n-1})-mv_n(y_1, \ldots, y_{n-1}), \quad \partial_i =
\frac{\partial}{\partial y_i}, \quad 1\leq i \leq n-1,
\end{gathered}
\right\}
\end{equation}
and where for an arbitrary polynomial $G(x_1,\ldots, x_n)$
we define
\[
g(y_1, \ldots, y_{n-1})\stackrel{\mathrm{ def}}{=}G(y_1 +
z_1, \ldots,
y_{n-1}+z_{n-1}, 1).
\]
As $z$ is a Darboux point, then all the polynomials $w_i$,
$1\leq i\leq
n-1$, vanish at the origin $y=0$.  Comparing the minimal
degree terms of
both sides in \myref{reduced} we obtained that
\[
   \sum_{i=1}^{n-1} l_i\frac{\partial h}{\partial y_i} =
\chi h,
\]
where
\[
     l_i = \sum_{j=1}^{n-1} l_{i j} y_j, \qquad 1 \leq i
\leq n-1,
\]
are homogeneous linear terms of $w_i$, $\chi$ is the zero
order term of $q$
and $h$ is the non-trivial homogeneous component of $f$ of
the lowest
degree.  Let us denote by  $\rho_i$, $ 1 \leq i \leq n-1 $
the eigenvalues
of  the matrix ${\bf L}= [l_{ij}]_{1\leq i,j\leq n-1}$.
Then, it can be
shown (see \cite[Lemma 2.3]{Moulin:95::} ) that if $F$ is
a Darboux
polynomial of \myref{gensys} then   there exist
non-negative integers $i_k$,
$1 \leq k \leq n-1$, such that
\begin{equation}
\label{maine}
\sum_{k=1}^{n-1} i_k\rho_k = \chi, \qquad
\sum_{k=1}^{n-1} i_k = \deg h \leq m.
\end{equation}
The eigenvalues of the matrix $\bf L$ we will call {\em
Lagutinskii-Levelt
exponents}. The proof of the non-integrability presented
here uses
essentially the relations \myref{maine}.

In the proof of our theorem we will also use the following
well known fact.
Let $ F\in \C[y]\backslash\C$ be a polynomial of a degree
$s$ in one
variable $y$. Let us denote by $y_i, 1\leq i\leq s$, the
roots of $F$ and by
$\alpha_i,  1\leq i\leq s$, their multiplicities
respectively.  As it is
well known, in this case
\begin{equation}
\label{sfrac}
\frac{F'}{F}= \sum_{i=1}^s\frac{\alpha_i}{y-y_i},
\end{equation}
where $F'$ is  the derivative of $F$ with respect to $y$.
Moreover, such
decomposition is unique. Indeed, if
\[
\frac{F'}{F}= \sum_{i=1}^p\frac{\beta_i}{y-y'_i},
\]
then $s=p$, $y_i=y'_i, 1\leq i\leq s$,     and
$\alpha_i=\beta_i, 1\leq
i\leq s$.

\section{Proof of the theorem}

The Halphen system is invariant with respect to
permutations of variables.
Let us denote by $\tau$ an  automorphism of  $\polyc$
induced by  a
permutation of $\{x_1,x_2,x_3\}$.  If by   $d_H$  we
denote the  derivation
defined by the Halphen system   then
\[
   \tau\circ d_H \circ \tau^{-1} = d_H.
\]
 From this invariance property of $d_H$ we have
immediately that
\begin{equation}
\label{lemma1}
\text{if}\quad  d_H(F) = P F\quad\text{then}\quad
d_H(\tau(F)) = \tau(P)
\tau(F).
\end{equation}

First we show the following
\begin{lemma}
\label{polint}
The Halphen system \myref{halphen} does not have any
polynomial first
integral.
\end{lemma}
\begin{proof}
Let us consider a Darboux point $z^0=(1,1,1)$   of the
Halphen system and
let us assume that $F\in\polyc\backslash\C$ is a
homogeneous first integral
of degree $m$ and $F\neq 0$ (see point (4) of sec. 2).
Direct calculations
show that for the chosen Darboux point we have (see
\myref{new})
\[
 w_1 =-y_1 (1 + 2 y_2 + y_1 y_2), \quad w_2 =-y_2 (1 + 2
y_2 + y_1
y_2),\quad q = m(1 - y_1y_2).
\]
Thus, the Lagutinskii-Levelt exponents are $\rho_1=\rho_2=
-1$, and $\chi =
m$.  Then (see \myref{maine})  there exist two
non-negative integers $i_1$
and $i_2$ such that $i_1\rho_1 +i_2\rho_2 = -(i_1 +i_2) =
m$ but $m>0$.
Contradiction.
\end{proof}
Although the Halphen system has no polynomial first
integrals it has
Darboux polynomials.  If we   denote
\begin{equation}
\label{dpall}
\left.
\begin{gathered}
  F_1 = x_1 - x_2, \qquad P_1 = -2 x_3, \\ F_2 = x_2 -
x_3, \qquad P_2 = -2
x_1, \\ F_3 = x_3 - x_1, \qquad P_3 = -2 x_2,
\end{gathered} \qquad\right\}
\end{equation}
then we have
\begin{equation*}
d_H(F_i) = P_i F_i, \qquad i=1,2,3.
\end{equation*}
In order to prove the theorem, we show that \myref{dpall}
are all
irreducible Darboux polynomials of the system.

For $f,g \in \cpoly $ we write $f\propto g$  if there
exists $a\in\C$,
$a\neq 0$,  such that $f=ag$. Note that every `eigenvalue'
$P$ for the
Halphen system must be linear, i.e., it has the form $P =
p_1 x_1 + p_2 x_2
+p_3 x_3$.

Now we show the following
\begin{lemma}

If $F\in\polyc\backslash\C$ is a  irreducible homogeneous
polynomial of
degree $m$ such that $d_H(F)=PF$, and $P=\alpha P_i$ for
some
$i\in\{1,2,3\}$, and $\alpha\in
\C$ then $F\propto F_i$ and $\alpha=1$.

\end{lemma}
\begin{proof}

First,  assume that $i=1$ and introduce new variables
\begin{equation*}
z_1 = x_1 - x_2, \qquad z_2 = x_2 - x_3,\qquad z_3 = x_1 +
x_3.
\end{equation*}
In the new variables the Halphen system has the form
\begin{equation*}
\dt{z_1} = z_1 l_1,\qquad \dt{z_2} = -z_2 l_2, \qquad
\dt{z_3} = \frac{1}{2}
l_1 l_2,
\end{equation*}
where
\begin{equation}
\label{defl}
l_1 = z_1 + z_2 - z_3, \qquad l_2= z_1 + z_2 + z_3.
\end{equation}
Darboux polynomials and their corresponding `eigenvalues'
\myref{dpall} in
terms of the new variables  have the form
\begin{equation}
\label{dpolz}
\left.
\begin{alignedat}{3}
F_1 &= z_1, &\qquad F_2&= z_2, &\qquad F_3 &= -z_1 -
z_2,\\ P_1 &= l_1,
&\qquad P_2&= -l_2,& \qquad P_3 &= l_3 = z_1 - z_2 -z_3.
\end{alignedat}\right\}
\end{equation}
Let us assume that there exists an irreducible homogeneous
Darboux
polynomial $F$ of degree $m\geq 1$ with the `eigenvalue' $
P=p_1z_1+p_2z_2+p_3z_3=\alpha l_1$ for some $\alpha\in\C$,
i.e.,
\begin{equation*}
z_1 l_1 \partial_1 F - z_2 l_2 \partial_2 F +
\frac{1}{2}l_1 l_2 \partial_3
F = P F;\qquad \partial_i =\frac{\partial}{\partial z_i},
\quad i=1,2,3.
\end{equation*}
Using the Euler identity for $F$, we can eliminate
$\partial_3 F$ from the
above equation and, taking into account \myref{defl},  we
obtain
\begin{equation}
\label{red}
z_1 l_1^2 \partial_1 F + z_2 l_2^2 \partial_2 F   = (m l_1
l_2 - 2 z_3 P)F.
\end{equation}
 We put $z_1=0$ in the above equation. The obtained
equation  has the form
\begin{equation}
\label{z10}
 z_2 \tilde{l}_2^2 \partial_2 \tilde{F}   = (m \tilde{l}_1
\tilde{l}_2 - 2
z_3 \tilde{P})\tilde{F},
\end{equation}
where we denote $\tilde{G}(z_2,z_3) = G(0, z_2,z_3)$ for
an arbitrary
polynomial $G$. Indeed, $\tilde{\partial_2 {F}}=
\partial_2 \tilde{F}$.  Now
suppose that $F\not\propto z_1$. Under this assumption
$\tilde{F} \neq 0$
because $F\neq 0$ is homogeneous and irreducible and thus
$\deg
\tilde{F}=\deg{F}=m$. From \myref{z10} we obtain
\begin{equation*}
\frac{\partial_2 \tilde{F}}{\tilde{F}}= -\frac{m + 2
p_3}{z_2} +
\frac{2(m+p_3)}{z_2+z_3} -
\frac{2(p_2-p_3)z_3}{(z_2+z_3)^2}.
\end{equation*}
For fixed $z_3\neq 0$ it represents decomposition
\myref{sfrac}, and, as a
consequence, we obtain that
\begin{equation}
\label{fin}
p_2-p_3=2\alpha=0,
\end{equation}
thus $P=0$. However, in such a case,   from  Lemma
\ref{polint} we know
that  $F$ is constant.  This  contradicts the  assumption
that $\deg F\geq
1$.  Consequently,  $F \propto z_1$ and thus $m=1$. This
ends the proof when
$i=1$. Permutational invariance of the Halphen  system
(see \myref{lemma1})
implies that the same is true when $i=2$ or $i=3$.
\end{proof}
Almost in the same way we prove the following
\begin{lemma}
If $F\in\polyc\backslash\C$ is an  irreducible homogeneous
polynomial   of
degree $m$ such that $d_H(F) = P F$, and $F\not\propto
F_1$,  $F\not\propto
F_2$ then $P=\alpha P_3$, $\alpha\in\C$.
\end{lemma}

\begin{proof}
Let us  take  $P = p_1 z_1 + p_2 z_2 + p_3 z_3$ in
equation \myref{red}.
 From the proof of the previous lemma we obtain that
$p_2=p_3$ (see
\myref{fin}).  Next,   put $z_2=0$ in equation
\myref{red}. We obtain that
\begin{equation}
\label{z20}
 z_1 \hat{l}_1^2 \partial_1 \hat{F}   = (m \hat{l}_1
\hat{l}_2 - 2 z_3
\hat{P})\hat{F},
\end{equation}
where now for an arbitrary polynomial $G$ we denote
$\hat{G} = G(z_1,
0,z_3)$.  Under our  assumption $\hat{F} \neq 0$ because
$F\neq 0$ is
homogeneous and irreducible and thus $\deg
\hat{F}=\deg{F}= m $. From
\myref{z20} we obtain
\begin{equation*}
\frac{\partial_1 \hat{F}}{\hat{F}}= -\frac{m + 2 p_3}{z_1}
+
\frac{2(m+p_3)}{z_1-z_3} -
\frac{2(p_1+p_3)z_3}{(z_1-z_3)^2}.
\end{equation*}
This implies that necessarily $p_1= -p_3$ and,
consequently,
\[
 P =-p_3(z_1 - z_2 -z_3)=-p_3 l_3.
\]
 But $l_3$ is equal to $P_3$ expressed  in variables
$(z_1, z_2, z_3)$ (see
\myref{dpolz}).
\end{proof}

\noindent
Now we can proceed to the proof of our theorem stated in
Introduction.

\noindent {\bf Proof.}
Let us assume that the Halphen system \myref{halphen}
possesses a rational
first integral $F=A/B$, with relatively prime
$A,B\in\polyc$.  From point
(1) in section 2 we know that polynomials $A$ and $B$ are
Darboux
polynomials corresponding to the same `eigenvalue'
$P\in\polyc$. From  Lemma
1  we deduce that $P\not\equiv 0$.  From  Lemmas  2 and 3
we deduce that the
only irreducible homogeneous Darboux polynomials of the
Halphen system are
those given by \myref{dpall}.  Let us observe that the
linear forms $P_i$,
$1\leq i\leq 3$ are linearly independent, and thus, from
the point
\myref{nomore} of the previous section, we deduce that all
the Darboux
polynomial of the Halphen system are  monoms $\alpha
F_1^{i_1}F_2^{i_2}F_3^{i_3}$  for some non-negative
integers $i_1$, $i_2$,
$i_3$ such that $i_1+i_2+i_3>0$,  and $\alpha\in\C$. It
follows that  both
$A$ and $B$ are of this form.  As they corresponds to the
same `eigenvalue'
$P$, by Lemmas 2 and 3 they proportional and thus they are
not relatively
prime as was assumed. Contradiction shows that our theorem
is true. $\Box$ %
\section*{Acknowledgments}
We sincerely thank  A.~Nowicki from Department of
Mathematics of N.
Copernicus University   for many inspiring discussions.  %
For the first
author this work was supported also by grant KBN 22149203.

The second author acknowledges the Institutes of Astronomy
and of
Mathematics of Toru\'n University for their hospitality
and excellent
working conditions during August 1994, when this work was
done.

\newcommand{\noopsort}[1]{}

\end{document}